\documentclass{article}
\usepackage{pdproc} 

\usepackage{graphicx}
\begin{document}
\title{IceCube: Neutrinos Associated with Cosmic Rays}
\author{Francis Halzen}
\address{Department of Physics, University of Wisconsin, Madison, WI 53706,USA
\\and\\
DESY, Zeuthen, Germany}

\abstract{After a brief review of the status of the
  kilometer-scale neutrino observatory IceCube, we discuss the
  prospect that such detectors discover the still-enigmatic sources of
  cosmic rays. After all, this aspiration set the scale of the
  instrument.  While only a ``smoking gun'' is missing for the
  case that the Galactic component of the cosmic-ray spectrum
  originates in supernova remnants, the origin of the extragalactic
  component remains as inscrutable as ever. We speculate on the role
  of the nearby active galaxies Centaurus\,A and M87.}


\section{The First Kilometer-Scale, High-Energy Neutrino Detector: IceCube}
\begin{figure}[here]
    \begin{center}
\includegraphics[width=0.5\textwidth]{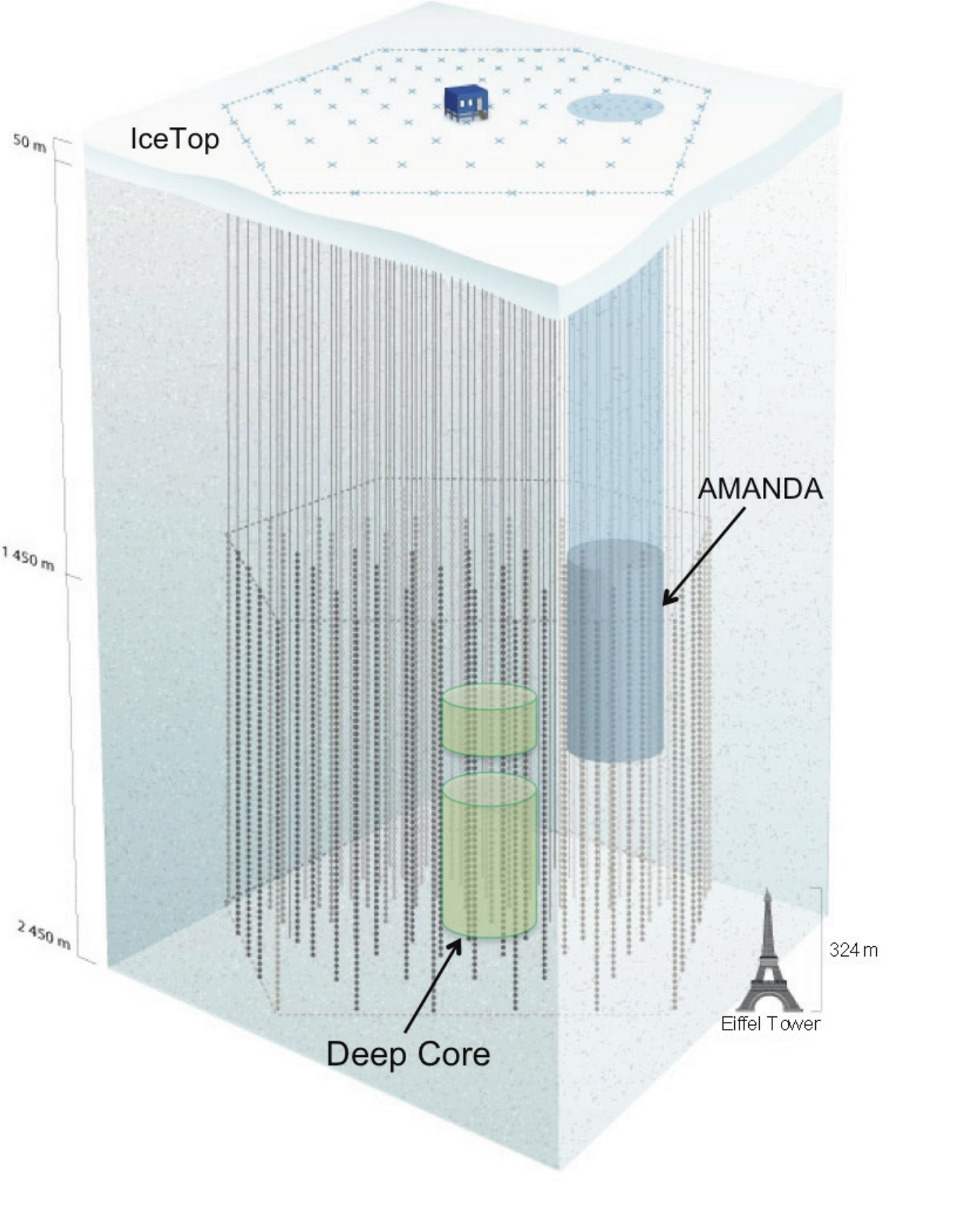}
    \end{center}
\caption {The IceCube detector, consisting of IceCube and IceTop 
and the low-energy sub-detector DeepCore. 
Also shown is the first-generation AMANDA detector.}
    \label{fig:deepcore}
 \end{figure}

\vspace{.2cm}
A series of first-generation experiments\cite{Spiering:2008ux} have
demonstrated that high-energy neutrinos with $\sim10$\,GeV energy and
above can be detected by observing Cherenkov radiation from
secondary particles produced in neutrino interactions inside large
volumes of highly transparent ice or water instrumented with a lattice
of photomultiplier tubes. The first second-generation detector,
IceCube  (see Fig.\ref{fig:deepcore}), is under construction at the geographic South
Pole\cite{Klein:2008px}. 

IceCube will consist of 80 km-length strings, each instrumented with 60 10-inch photomultipliers spaced by
17\,m. The deepest module is located at a depth of 2.450\,km so that
the instrument is shielded from the large background of cosmic rays at
the surface by approximately 1.5\,km of ice. Strings are arranged
at apexes of equilateral triangles that are 125\,m on a side. The
instrumented detector volume is a cubic kilometer of dark, highly
transparent and sterile Antarctic ice. Radioactive
background is dominated by instrumentation deployed into this
natural ice.

Each optical sensor consists of a glass sphere containing the
photomultiplier and the electronics board that digitizes the signals
locally using an on-board computer. The digitized signals are given a
global time stamp with residuals accurate to less than 3\,ns and are
subsequently transmitted to the surface. Processors at the surface
continuously collect these time-stamped signals from the optical
modules; each functions independently.  The digital messages are sent
to a string processor and a global event trigger. They are
subsequently sorted into the Cherenkov patterns emitted by secondary
muon tracks that reveal the direction of the parent
neutrino\cite{Halzen:2006mq}.

\begin{figure}[here]
\begin{center}
\includegraphics[width=0.5\textwidth]{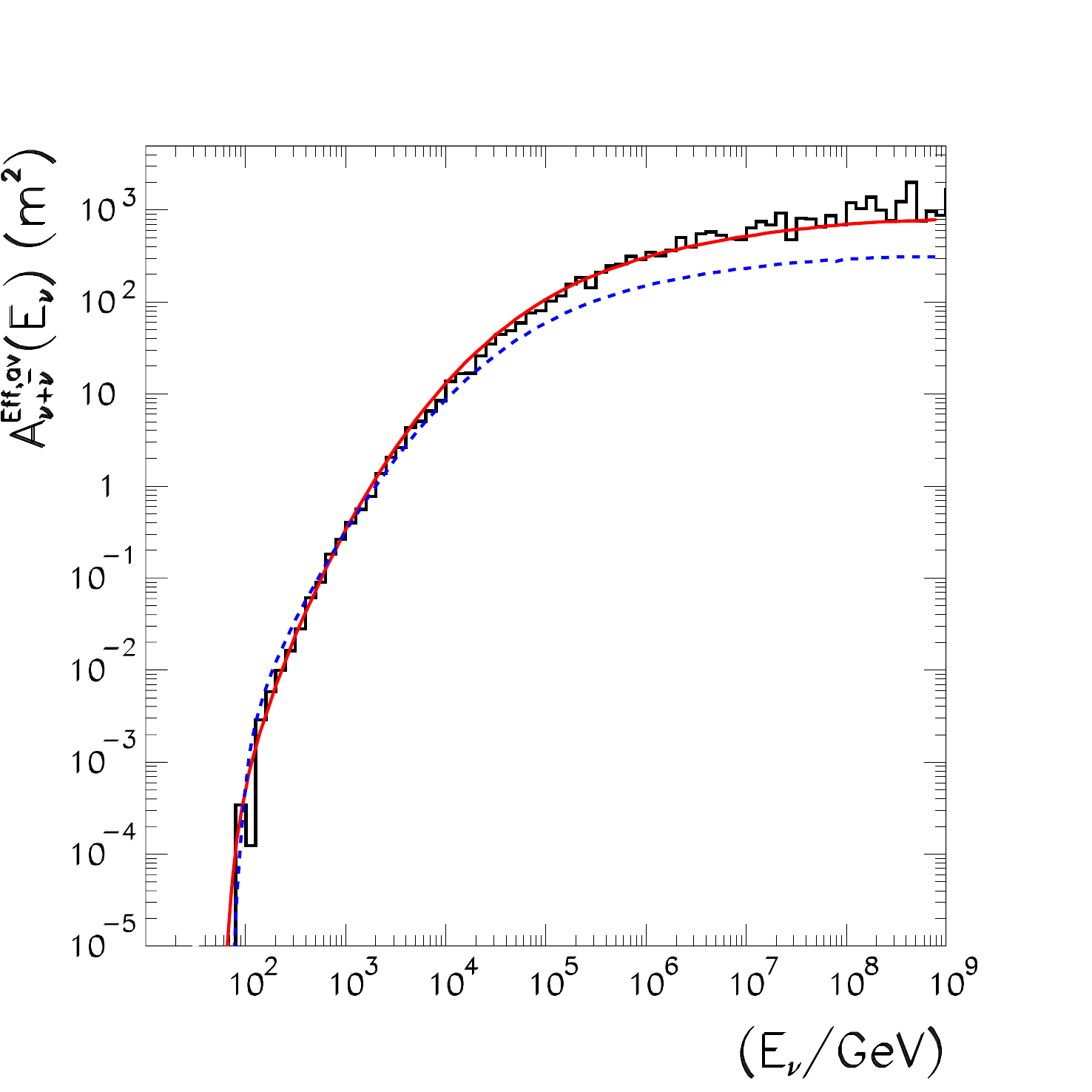}
\end{center}
\caption{The neutrino effective area (averaged over the Northern
Hemisphere) from IceCube simulation (black histogram) is compared to
the convolution of the approximate muon effective area from
reference\protect\cite{GonzalezGarcia:2009jc} (solid red line) that we will
use in the various estimates of event rates throughout this paper. The
neutrino area exceeds the design area (shown as the dashed blue line) \protect\cite{ic2004}
at high energy. }
\label{fig:areas}
\end{figure}

Based on data taken with 40 of the 59 strings that have already been deployed, the anticipated effective area of the completed IceCube detector is shown in Fig.\ref{fig:areas}. Notice the factor 2 to 3 increase in effective area over what had been anticipated\cite{ic2004}.

Despite its discovery potential touching a wide range of scientific issues, construction of IceCube has been largely motivated by the possibility of opening a new window on the Universe using neutrinos as cosmic messengers. Specifically, we will revisit IceCube's prospects to detect cosmic neutrinos associated with cosmic rays and to reveal their sources prior to the 100th anniversary of their discovery by Victor Hess in 1912.

\begin{figure}
\begin{center}
\includegraphics[width=0.6\textwidth]{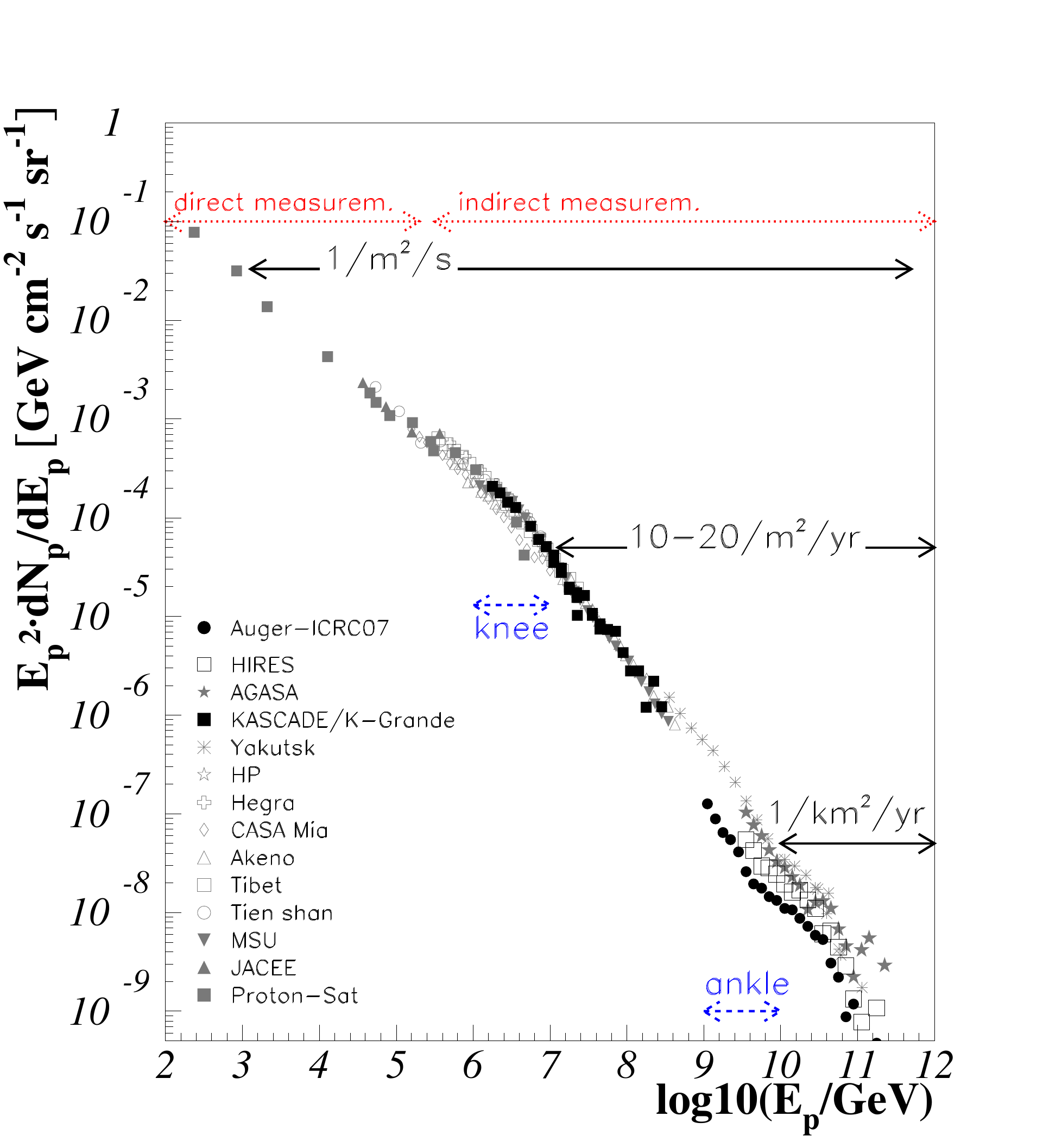}
\end{center}
\caption{At the energies of interest here, the cosmic-ray spectrum
follows a sequence of 3 power laws. The first 2 are separated by the
``knee'', the 2nd and 3rd by the ``ankle''. Cosmic
rays beyond the ankle are a new population of particles produced in
extragalactic sources.}
\label{fig:fig02}
\end{figure}

Cosmic accelerators produce particles with energies in excess of $10^8$\,TeV; we still do not know where or how\cite{Sommers:2008ji}. The flux of cosmic rays observed at Earth is shown in Fig.\ref{fig:fig02}. The energy spectrum follows a sequence of three power laws. The first two are separated by a feature dubbed the ``knee'' at an energy\footnote{We will use energy units TeV, PeV and EeV, increasing by factors of 1000 from GeV energy.} of approximately 3\,PeV.  There is evidence that cosmic rays up to this energy are Galactic in origin.  Any association with our Galaxy disappears in the vicinity of a second feature in the spectrum referred to as the ``ankle"; see Fig.\ref{fig:fig02}. Above the ankle, the gyroradius of a proton in the Galactic magnetic field exceeds the size of the Galaxy, and we are witnessing the onset of an extragalactic component in the spectrum that extends to energies beyond 100\,EeV. Direct support for this assumption now comes from two experiments \cite{Abraham:2008ru} that have observed the telltale structure in the cosmic-ray spectrum resulting from the absorption of the particle flux by the microwave background, the so-called Greissen-Zatsepin-Kuzmin cutoff. The origin of the flux in the intermediate region covering PeV-to-EeV energies remains a mystery, although it is routinely assumed that it results from some high-energy extension of the reach of Galactic accelerators.

Acceleration of protons (or nuclei) to TeV energy and above requires massive bulk flows of relativistic charged particles. These are likely to originate from exceptional gravitational forces in the vicinity of black holes or neutron stars. Gravity powers large currents of charged particles that are the origin of high magnetic fields. These create the opportunity for particle acceleration by shocks. It is a fact that electrons are accelerated to high energy near black holes; astronomers detect them indirectly by their synchrotron radiation. Some must accelerate protons because we observe them as cosmic rays.

How many gamma rays and neutrinos are produced in association with the cosmic-ray beam?\footnote{We do not discuss cosmic neutrinos directly produced in the interactions of cosmic rays with microwave photons here; extensions of IceCube have been proposed to detect them\cite{Allison:2009rz}.} Generically, a cosmic-ray source should also be a beam dump. Cosmic rays accelerated in regions of high magnetic fields near black holes inevitably interact with radiation surrounding them, e.g., UV photons in active galaxies or MeV photons in gamma-ray--burst fireballs. In these interactions, neutral and charged pion secondaries are produced by the processes
\begin{eqnarray*}
p + \gamma \rightarrow \Delta^+ \rightarrow \pi^0 + p
\mbox{ \ and \ }
p + \gamma \rightarrow \Delta^+ \rightarrow \pi^+ + n.
\end{eqnarray*}
While secondary protons may remain trapped in the high magnetic fields, neutrons and the decay products of neutral and charged pions escape. The energy escaping the source is therefore distributed among cosmic rays, gamma rays and neutrinos produced by the decay of neutrons, neutral pions and charged pions, respectively. In the case of Galactic supernova shocks, cosmic rays interact with gas, e.g. with dense molecular clouds, as well as radiation,
producing equal numbers of pions of all three charges in hadronic collisions $p+p \rightarrow n\,[\,\pi^{0}+\pi^{+} +\pi^{-}]+X$.

Kilometer-scale neutrino detectors have the sensitivity to reveal generic cosmic-ray sources with an energy density in neutrinos comparable to their energy density in cosmic rays\cite{TKG} and pionic TeV gamma rays\cite{AlvarezMuniz:2002tn}.

\section{Sources of Galactic Cosmic Rays}
\vspace{.2cm}

Supernova remnants were proposed as possible sources of Galactic cosmic rays as early as 1934 by Baade and Zwicky\cite{zwicky}; their proposal is still a matter of debate after more than 70 years.  Galactic cosmic rays reach energies of at least several PeV, the ``knee" in the spectrum. Their interactions with Galactic hydrogen in the vicinity of the accelerator should generate gamma rays from decay of secondary pions that reach energies of hundreds of TeV. Such sources should be identifiable by a relatively flat energy spectrum that extends to hundreds of TeV without attenuation; they have been dubbed PeVatrons. Straightforward energetics arguments are sufficient to conclude that present air Cherenkov telescopes should have the sensitivity necessary to detect TeV photons from PeVatrons\cite{GonzalezGarcia:2009jc}.

They may have been revealed by an all-sky survey in $\sim 10$\,TeV gamma rays with the Milagro detector\cite{Abdo:2006fq}. A subset of sources located within nearby star-forming regions in Cygnus and in the vicinity of Galactic latitude $l=40$\,degrees are identified; some are not readily associated with known supernova remnants or with non-thermal sources observed at other wavelengths.  Subsequently, directional air Cherenkov telescopes were pointed at three of the sources, revealing them as PeVatron candidates with an approximate $E^{-2}$ energy spectrum that extends to tens of TeV without evidence for a cutoff\, \cite{hesshotspot,magic2032}, in contrast with the best studied supernova remnants RX J1713-3946 and RX J0852.0-4622 (Vela Junior).

Some Milagro sources may actually be molecular clouds illuminated by the cosmic-ray beam accelerated in young remnants located within $\sim100$\,pc. One expects indeed that multi-PeV cosmic rays are accelerated only over a short time period, when the remnant transitions from free expansion to the beginning of the Sedov phase and the shock velocity is high. The high-energy particles can produce photons and neutrinos over much longer periods when they diffuse through the interstellar medium to interact with nearby molecular clouds; for a detailed discussion, see reference~\cite{gabici}. An association of molecular clouds and supernova remnants is expected in star-forming regions.

\begin{figure}[h]
\centering
\includegraphics[width=4in]{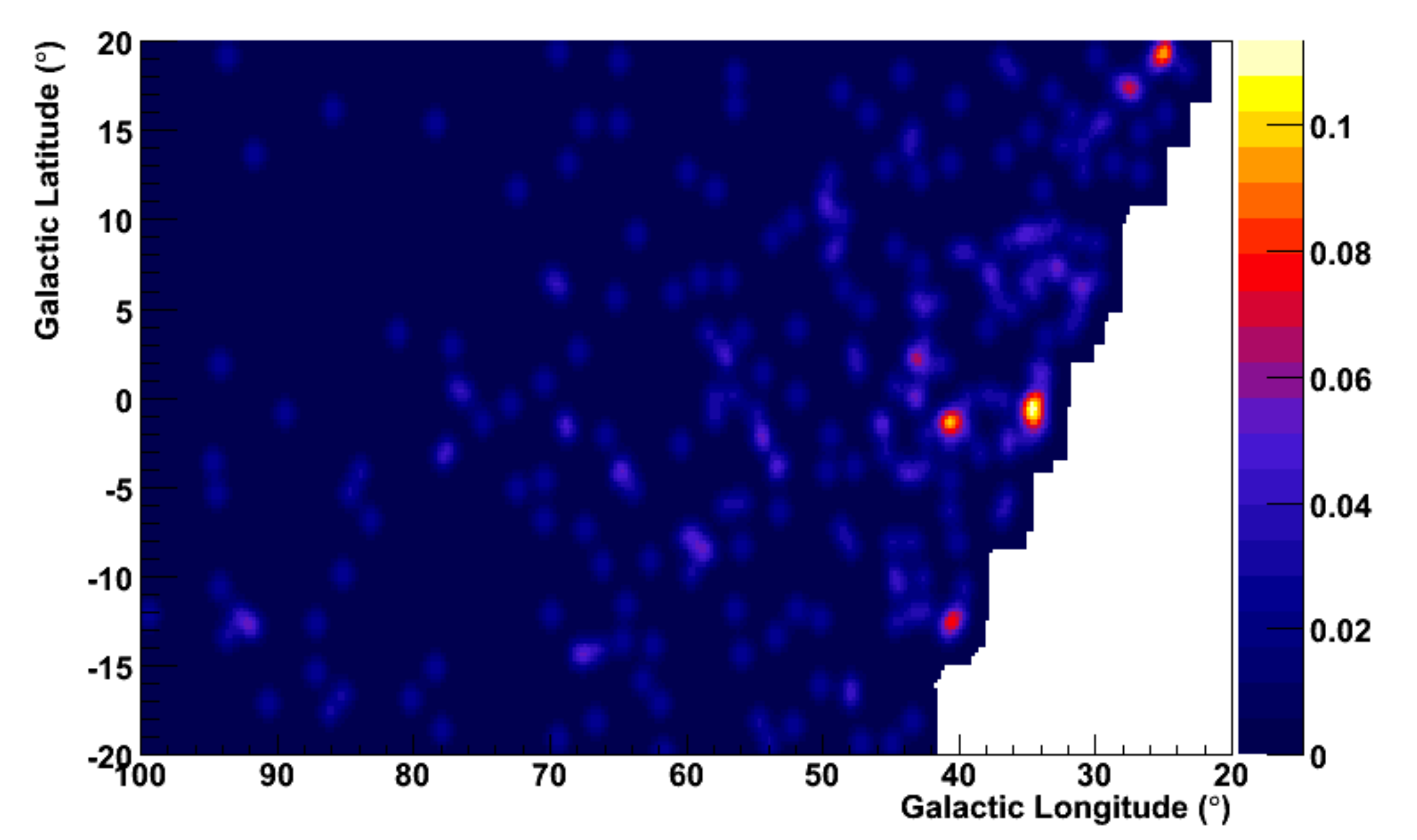}
\caption{Simulated sky map of IceCube in Galactic coordinates after 5
years of operation of the completed detector.  Two Milagro
sources are visible ``by eye" with 4 events for MGRO J1852+01 and 3
events for MGRO J1908+06 with energy in excess of 40\,TeV. These, as
well as the background events, have been randomly distributed according
to the resolution of the detector and the size of the sources.}
\label{fig:5year_Map_1}
\end{figure}

Despite the rapid development of instruments with improved sensitivity, it has been impossible to conclusively pinpoint supernova remnants as the sources of cosmic rays by identifying accompanying gamma rays of pion origin. Detecting the accompanying neutrinos would provide incontrovertible evidence for cosmic ray acceleration in the sources. Particle physics dictates the relation between gamma rays and neutrinos and basically predicts the production of a $\nu_\mu+\bar\nu_\mu$ pair for every two gamma rays seen by Milagro. This calculation can be performed in a more sophisticated way with approximately the same outcome. We conclude that, within uncertainties in the source parameters and the detector performance, confirmation that Milagro mapped sources of Galactic cosmic rays should emerge after operating the complete IceCube detector for several years; see Fig.\ref{fig:5year_Map_1}.

The quantitative statistics can be summarized as follows. For average values of the parameters describing the flux, we find that the completed IceCube detector could confirm sources in the Milagro sky map as sites of cosmic-ray acceleration at the
$3\sigma$ level in less than one year and at the $5\sigma$ level in three years\cite{GonzalezGarcia:2009jc}. These results agree with previous estimates \cite{hkm}. There are intrinsic ambiguities in this estimate. With the performance of IceCube now well understood, these are of astrophysical origin. In the absence of observation of TeV-energy supernova neutrinos by IceCube, the nature of sources that produce cosmic rays near the knee of the spectrum remains unresolved.

\section{Sources of Extragalactic Cosmic Rays}
\vspace{.2cm}

Although there is no direct evidence that supernovae accelerate cosmic rays, the idea is generally accepted because of energetics: three supernovae per century converting a reasonable fraction of a solar mass into particle acceleration can accommodate the steady flux of cosmic rays in the Galaxy. Energetics also drive speculation on the origin of extragalactic cosmic rays.

By integrating the cosmic-ray spectrum in Fig.\ref{fig:fig02} above the ankle, we find that the energy density of the Universe in extragalactic cosmic rays is $\sim 3 \times10^{-19}\rm\,erg\ cm^{-3}$\cite{TKG}. The power required for a population of sources to generate this energy density over the Hubble time of $10^{10}$\,years is $\sim 3 \times 10^{37}\rm\,erg\ s^{-1}$ per (Mpc)$^3$. A gamma-ray--burst (GRB) fireball converts a fraction of a solar mass into the acceleration of electrons, seen as synchrotron photons. The energy in extragalactic cosmic rays can be accommodated with the reasonable assumption that shocks in the expanding GRB fireball convert roughly equal energy into the acceleration of electrons and cosmic rays\cite{waxmanbahcall}. It so happens that $\sim 2 \times 10^{52}$\,erg per GRB will yield the observed energy density in cosmic rays after $10^{10}$ years given that the rate is of order 300 per $\textrm{Gpc}^{3}$ per year. Hundreds of bursts per year over Hubble time produce the observed cosmic ray density, just like 3 supernova per century accommodate the steady flux in the Galaxy.

\begin{figure}[h]
\centering
\includegraphics[width=3in]{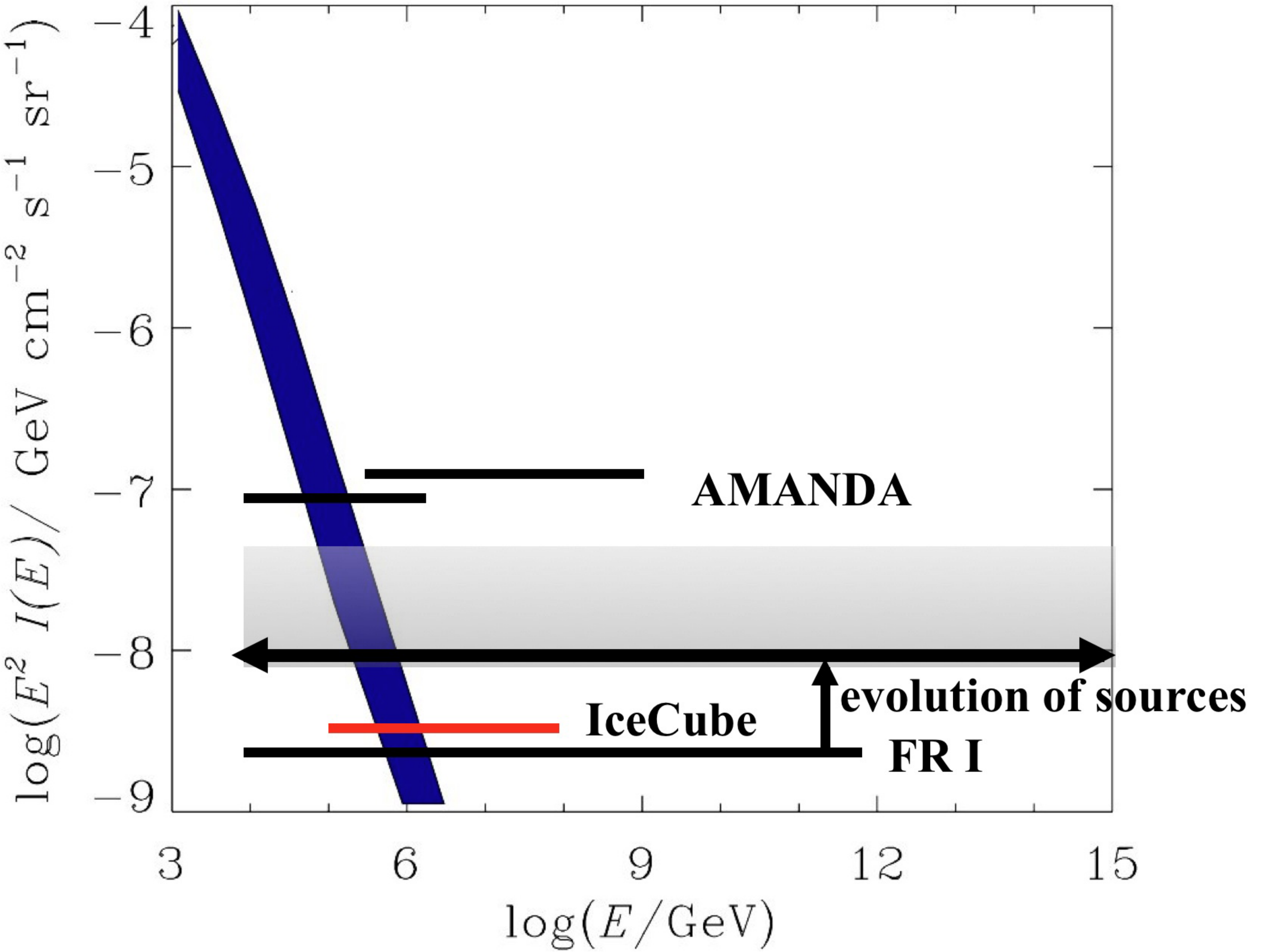}
\caption{Our energetics estimate of the flux of neutrinos associated
with sources of the highest-energy cosmic rays (the shaded range)
is compared to limits established by AMANDA and the
sensitivity of IceCube \protect\cite{merida}. Also shown is the flux derived
from the assumption that AGN are the sources which we model using
spectral energy information on nearby active galaxies Cen A and
M87; see Fig.\ref{fig:cen_A_sed}. Integration of AGN to larger
redshifts and evolution of sources may reconcile the two estimates which differ by a factor of
3. Also shown is the background flux of atmospheric neutrinos at lower energies.}
\label{fig:agn_spectrum}
\end{figure}

Problem solved? Not really, it turns out that the same result can be achieved assuming that active galactic nuclei (AGN) convert $\sim 2 \times 10^{44}\rm\,erg\ s^{-1}$ per AGN into particle acceleration. As is the case for GRB, this is an amount that matches their output in electromagnetic radiation. Whether GRB or AGN, the observation that these sources radiate similar energies in photons and cosmic rays is consistent with the beam dump scenario previously discussed. In the interaction of cosmic rays with radiation and gases near the black hole, roughly equal energy goes into the secondary neutrons and neutral pions whose energy ends up in cosmic rays and gamma rays, respectively. It predicts a matching flux of neutrinos that, after many correction factors that cancel, is predicted to be roughly equal to the cosmic ray flux:
\begin{equation}
E_{\nu}^{2} dN / dE_{\nu}= 5 \times 10^{-8}\rm\,GeV \,cm^{-2}\, s^{-1}\, sr^{-1}
\label{extragalactic}
\end{equation}
This is referred to as the Waxman-Bahcall ``bound" \cite{WB}. If we adjust it downward by a factor 5 because only 20\% of the proton energy is transferred to pions, we obtain a generic neutrino flux predicted by the GRB and AGN scenarios. After seven years of operation, AMANDA's sensitivity is approaching the interesting range, but it takes IceCube to explore it; see Fig.\ref{fig:agn_spectrum}.

If AGN are indeed the sources, the proximity of the Fanaroff-Riley I (FRI) active galaxies Cen A and M87 singles them out as potential accelerators\cite{anchordoquicena}. As we did for the Milagro sources, we will attempt to translate their TeV gamma ray into a neutrino flux\, although interpreting TeV gamma-ray observations is challenging.  The high-energy emission of AGN is extremely variable, and it is difficult to compare multi-wavelength data taken at different times. Our best guess is captured in Fig.\ref{fig:cen_A_sed} where the TeV flux is shown along with observations of the multi-wavelength emission of Cen A compiled by Lipari\cite{lipari}.

The TeV flux shown represents an envelope of observations:
\begin{enumerate}
\item Archival data of TeV emission of Cen A collected in the early 1970s, with the Narrabri optical intensity interferometer of the University of Sydney\cite{cenasydney}.
\item Observation by HEGRA\cite{hegra} of M87. We scaled the flux of M87 at 16\,Mpc to the distance to Cen A. After adjusting for the different thresholds of the HEGRA and Sydney experiments, we find identical source luminosities for M87 and Cen A of roughly $7\times10^{40}\, {\rm erg\,s^{-1}}$, assuming an $E^{-2}$ gamma-ray spectrum.
\item The time-averaged gamma-ray flux thus obtained is close to the flux from Cen A recently observed by the H.E.S.S. collaboration\cite{Aharonian:2009xn}.
\end{enumerate}

Given that we obtain identical luminosities for Cen A and M87, we venture the assumption that they may be generic FRI, a fact we will exploit to construct the diffuse neutrino flux from all FRI. The conversion of TeV gamma rays to neutrinos, exploited for Galactic sources, yields a neutrino flux for a single generic FRI of
\begin{equation}
\frac{dN_{\nu}}{dE} \simeq 5\times10^{-13} \left( \frac{E}{\rm TeV}
\right)^{-2} {\rm TeV^{-1}\,cm^{-2}\,s^{-1},}
\end{equation}
Whereas the assumption that TeV gamma rays are of pionic origin is an educated guess in the case of supernovae, here it is merely an assumption that can be tested by neutrino observations.

The total diffuse flux from all such sources with a density of $n \simeq 8\times 10^{4} \,{\rm Gpc^{-3}}$ within a horizon of $R\sim3 \,{\rm Gpc}$\cite{fridensity} is simply the sum of luminosities of the sources weighted by their distance, or
\begin{equation}
\frac{dN_\nu}{dE_{\rm diff}} = \sum \frac{L_{\nu}}{4\pi d^{2}} =L_{\nu} \, n\,R = 4\pi d^{2} n R\frac{dN_\nu}{dE},
\end{equation}
where $dN_\nu/dE$ is given by the single source flux. We performed the sum by assuming that the galaxies are uniformly distributed. This evaluates to:
\begin{equation}
\frac{dN_\nu}{dE_{\rm diff}} = 2\times 10^{-9}\,\left(\frac{E}{\rm GeV}\right)^{-2}\,{\rm GeV^{-1}\,cm^{-2}\,s^{-1}\,sr^{-1}},
\end{equation}
approximately a factor 3 below the flux previously estimated on the basis of source energetics; see Fig.\ref{fig:cen_A_sed}.

\begin{figure}[here]
\centering\includegraphics[width=4in]{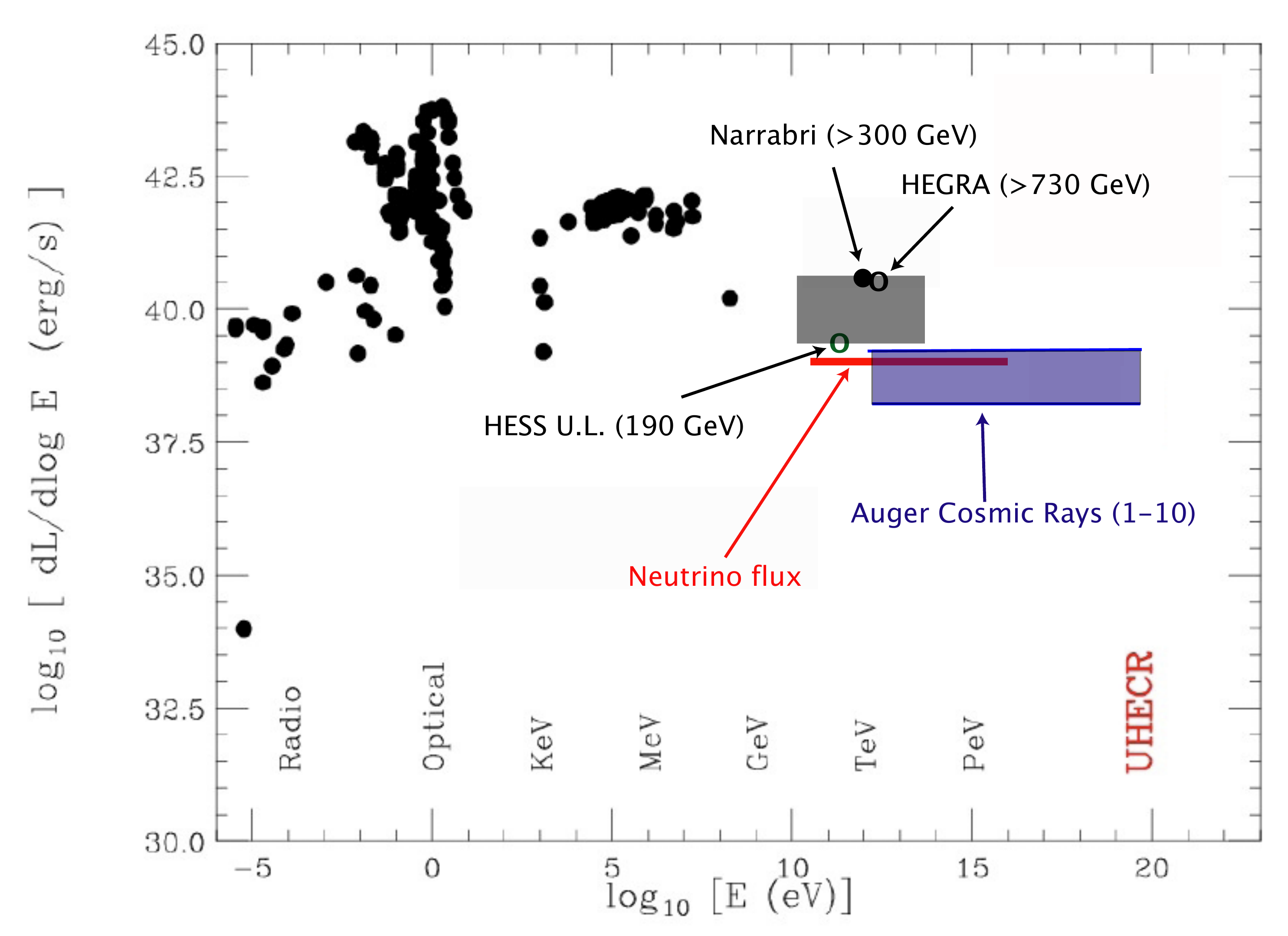}
\caption{Spectral energy distribution of Cen A (black dots). Keeping in
mind that the source is variable, we show our estimates for the flux
of TeV gamma rays (upper gray shading) and cosmic rays assuming that between
1 and 10 events observed by Auger originated at Cen A (lower blue
shading). We note that cosmic-ray and TeV gamma-ray fluxes
estimated in this paper are at a level of the electromagnetic
component shown from radio waves to GeV photons. Our estimate for the
neutrino flux (labeled ``Neutrino flux") is shown as the red line. }
\label{fig:Cen_A_SED}
\end{figure}

The neutrino flux from a single source such as Cen A is clearly small: repeating the calculation for power-law spectra between 2.0 and 3.0, we obtain, in a generic neutrino detector of effective muon area $1\,{\rm km^{2}}$, only 0.8 to 0.02 events per year only. The diffuse flux yields a more comfortable event rate of between 19 and 0.5 neutrinos per year. Considering sources out to 3\,Gpc, or a redshift of order 0.5 only, is probably conservative. Extending the sources beyond $z\sim1$, and taking into account their possible evolution, may increase the flux by a factor 3 or so. We might thus bridge the gap between this and the previous estimate of the cosmic-neutrino flux based solely on energetics of the cosmic rays.

Finally, having previously argued for a relation between the energetics of TeV gamma rays, cosmic rays and neutrinos, we ask the question what Auger data may reveal about the flux of the source? The answer depends on the number of events in their sky map\cite{auger} that actually correlate with Cen A. The range of cosmic-ray fluxes shown in Fig.\ref{fig:Cen_A_SED} corresponds to 1--10 events out of a total of 26.

If it turns out that GRB are the sources, the task is simplified.  Should this be the case for cosmic rays, we expect to observe just under 10 neutrinos per year\cite{guetta} in coincidence with Swift and Fermi GRB, which corresponds to a $5\sigma$ observation.  A similar statistical power can be obtained by detecting showers produced by electron and tau neutrinos.

In summary, while the road to identification of the sources of the Galactic cosmic ray has been mapped, the origin of the extragalactic component remains as mysterious as ever.

A more detailed version of this talk can be found in the proceedings of the  "XIII International Workshop on Neutrino Telescopes",  "Istituto Veneto di Scienze, Lettere ed Arti", Venice, Italy (2009).

\section{Acknowledgements} I would like to thank my collaborators Elisa Bernardini, Concha Gonzalez-Garcia, Darren Grant and Alexander Kappes, as well as John Beacom, Julia Becker, Peter Biermann, Steen Hannestad, Christian Spiering and Stefan Westerhoff for valuable discussions. This research was supported in part by the U.S. National Science Foundation under Grants No.~OPP-0236449 and  PHY-0354776; by the U.S.~Department of Energy
under Grant No.~DE-FG02-95ER40896; by the University of Wisconsin Research Committee with funds granted by the Wisconsin Alumni Research Foundation; and by the Alexander von Humboldt Foundation in Germany.

\end{document}